\documentclass{PoS}
\usepackage[utf8]{inputenc}
\usepackage{cite}
\title{Commissioning and first results from the CMS phase~1 upgrade pixel detector}

\ShortTitle{The CMS phase~1 upgrade pixel detector}

\author{\speaker{Jory Sonneveld on behalf of the CMS collaboration}\\%\thanks{A footnote may follow.}\\
        Hamburg University\\
        E-mail: \email{jory.sonneveld@cern.ch}}

\abstract{

    The phase~1 upgrade of the CMS pixel detector has been designed to maintain the tracking performance at instantaneous luminosities of $2 \times 10^{34} \mathrm{~cm}^{-2} \mathrm{~s}^{-1}$.
    Both barrel and endcap disk systems now feature one extra layer (4 barrel layers and 3 endcap disks), and a digital readout that provides a large enough bandwidth to read out its 124M pixel channels (87.7 percent more pixels compared to the previous system).
    The backend control and readout systems have been upgraded accordingly from VME-based to micro-TCA-based ones.
    The detector is now also equipped with a bi-phase CO$_2$ cooling system that reduces the material budget in the tracking region. The detector has been installed inside CMS at the start of 2017 and is now taking data.
    These proceedings discuss experiences in the commissioning and operation of the CMS phase~1 pixel detector.
    The first results from the CMS phase~1 pixel detector with this year's LHC proton-proton collision data are presented.
    The new pixel detector outperforms the previous one in terms of hit resolution, tracking, and vertex resolution.

}

\FullConference{The 26th International Workshop on Vertex Detectors\\
		10-15 September, 2017\\
		Las Caldas, Asturias, Spain}

\begin{document}

\section{The CMS phase~1 pixel detector: a necessary replacement}
%At CERN in Geneva, the Large Hadron Collider (LHC) experiment collides protons with 6.5 TeV per proton beam and ideally 2808 bunches per beam at 25ns bunch spacing (40MHz).
The Compact Muon Solenoid (CMS)
\cite{Chatrchyan:2008aa}
is a general-purpose detector designed to study proton-proton and heavy-ion  collisions at the Large Hadron Collider (LHC) of CERN. % with a design luminosity of
%$2\times 10^{34}\mathrm{cm}^{-2}\mathrm{s}^{-1}$.
%Erik: difficult, phase~0 already went to 1e34.
The phase 1 detector features a strong solenoidal magnetic field of 3.8 Tesla ensuring good momentum  resolution  for  charged  particles.
In the heart of CMS, within a silicon  outer  tracker, particle trajectories are recorded by a vertex detector using pixel sensors of cell size $100 \times 150$ $\mu$m$^2$.
Until 2016, the so-called phase~0 pixel detector covered an area of 1.06 m$^2$ \cite[p34]{Chatrchyan:2008aa} with 66 million channels until it was replaced by the CMS phase~1 pixel detector
\cite{CMS:2012sda}
that has 124 million pixel channels, or 87.7\% more pixels compared to the previous system.
The phase~1 pixel detector consists of n+~in~n sensors with 66560 pixels of the same size as in the phase 0 system 100 $\times$ 150 $\mu$m$^2$.
Each sensor covers a total active area of $16.2 \times 64.8$ mm$^2$ per 16 readout chips (one module), so that all 1856 modules of the new pixel detector cover 1.95 m$^2$.

\begin{figure}
    \centering
    \includegraphics[width=0.35\textwidth]{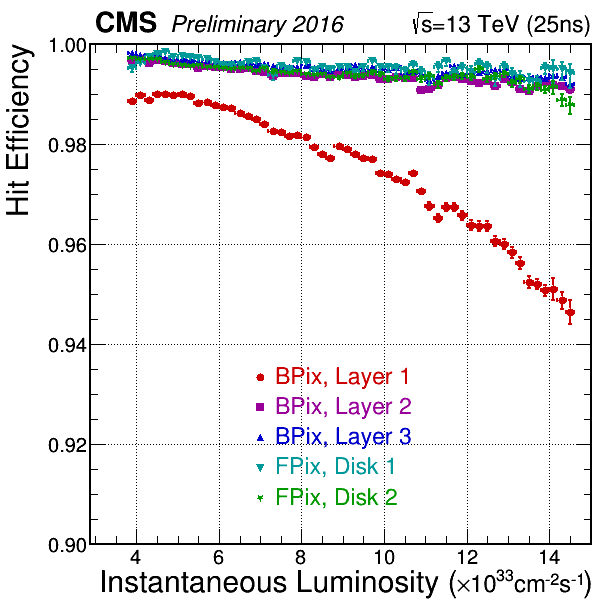}
    \includegraphics[width=0.35\textwidth]{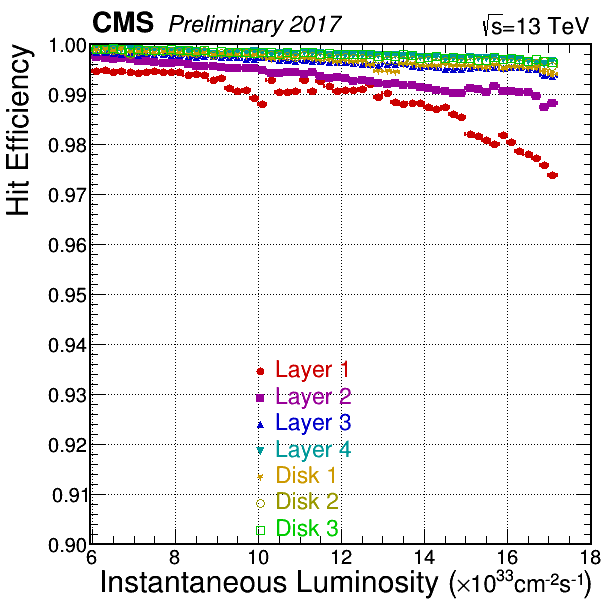}
    \caption{Hit efficiency as a function of instantaneous luminosity for the phase~0 (left) and phase~1 (right) pixel detector. Figures from \cite{pixeloffline2016,cmstracker}.}
    \label{fig:hiteff}
\end{figure}

The forward pixel detector now features an additional disk, and the barrel pixel detector an additional layer compared to the phase~0 pixel detector.
Even with the same geometry as the new detector, however, the phase~0 pixel detector would not have been able to cope with the rates that we observe at the LHC today, which are
well in excess of
%up to twice its design luminosity of
$1\times 10^{34}\mathrm{cm}^{-2}\mathrm{s}^{-1}$.
%compared to the peak luminosity of
%$1.53\times 10^{33}\mathrm{cm}^{-2}\mathrm{s}^{-1}$
This can be seen in Fig.~\ref{fig:hiteff}, where the hit efficiency for layer 1 in the phase 0 detector sharply decreases for higher instantaneous luminosity, while it does not for the current pixel detector.
For this reason it was necessary to replace the phase~0 pixel detector in the Extended Year-End Technical Stop (EYETS) 2016/2017.

The readout for the phase 1 detector is improved accordingly to 400~Mbit/s digital, which is less prone to error and has better rates compared to the 40~MHz analog that was available before.
The backend now features micro-TCA (Telecommunications Computing Architecture) compared to the VME(Versa Module Europa)-based architecture from before, and a reduced material budget was achieved with a two-phase CO$_2$ cooling system
\cite{Tropea:2016jym}
compared to the single-phase C$_6$F$_{14}$ cooling used for the previous pixel detector.

The innermost layer of the current pixel detector is designed to last until the long shutdown~2 that takes place in 2019 and 2020. At this point, layer 1 of the pixel detector is to be replaced because of radiation damage. Thereafter the same detector is to be used in data taking until long shutdown~3 when the high-luminosity phase~2 detector is to be installed.
% rates
% four layers
% timeline: up to ls2
% l1 replacement
\section{Phase~1 pixel detector performance}
\begin{figure}
    \centering
    \includegraphics[width=0.28\textwidth]{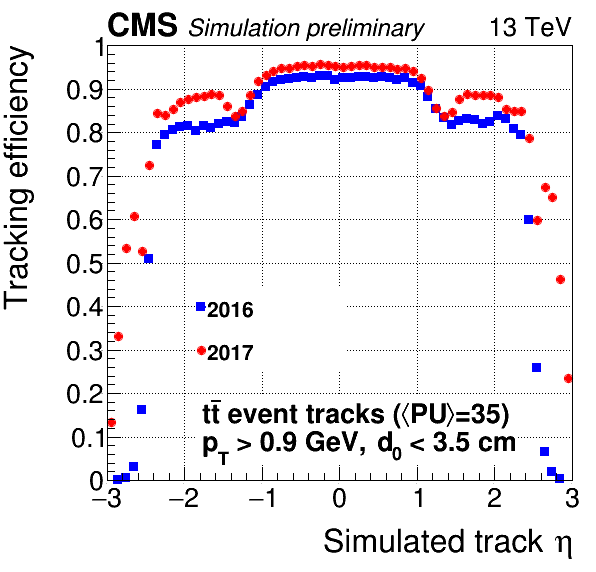}
    \includegraphics[width=0.28\textwidth]{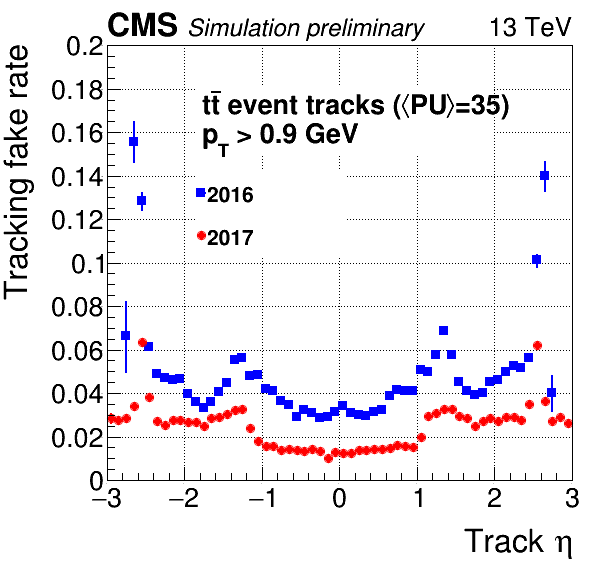}
    \includegraphics[width=0.36\textwidth, trim=250 580 250 540, clip]{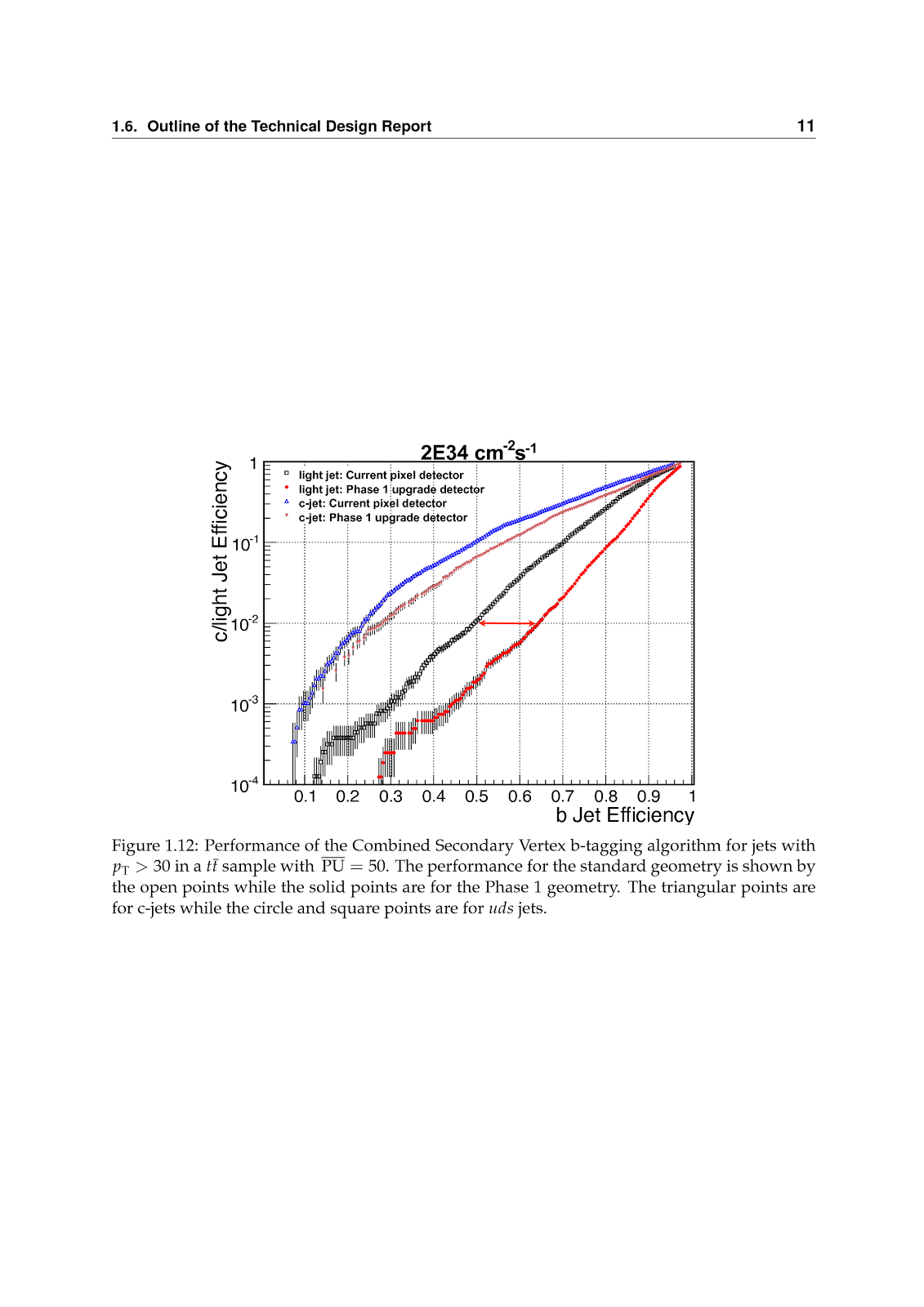}
    \caption{Tracking efficiency, fake rate and b-mistag rate as a function of b-tag efficiency (simulation) for the phase~0 (blue, black) and phase~1 (red) pixel detectors. Figures from \cite{CMS:2012sda,trackerperformance}.}
    \label{fig:trackingeffsimulation}
\end{figure}

%\paragraph{Tracking performance}
%The CMS phase~1 pixel detector is expected to keep a much higher hit efficiency with the increased luminosity of the LHC in 2017 compared to 2016, as shown in Fig.~\ref{fig:hiteff}.
In addition to a higher hit efficiency, the phase~1 pixel detector is also expected to contribute to a higher tracking efficiency and lower tracking fake rate in the entire coverage in pseudorapidity; see the efficiency and fake track rate for simulated $t\bar{t}$ events in Fig.~\ref{fig:trackingeffsimulation}.
A fake track is a reconstructed track not matched to a truth track in simulation \cite[p20]{CMS:2012sda}.
Pseudorapidity is defined as
%In this coordinate system the z axis points along the anticlockwise beam direction, the x axis points towards the center of the ring, and the y axis points upwards. The polar angle is defined with respect to the positive z axis, as shown in this figure:
%\footnote{
%The pseudorapidity $\eta$ is defined as $\eta = -\ln(\tan(\theta/2))$, with $\theta$ the polar angle in the CMS coordinate system. In this coordinate system the z axis points along the anticlockwise beam direction, the x axis points towards the center of the ring, and the y axis points upwards. The polar angle is defined with respect to the positive z axis, as shown in this figure:
% \includegraphics[width=0.32\textwidth]{img/Figures_T_Coordinate.png} (figure from \cite{Schott:2014sea}).}
$\eta = -\ln(\tan(\theta/2))$, where $\theta$ is the polar angle in the CMS coordinate system.
%$\eta$, see Fig.~\ref{fig:trackingeffsimulation}.
The b-jet tagging efficiency, too, is expected to improve tremendously with the new pixel detector, as shown with the mistag rate as a function of b-tag efficiency on the right in the same figure.
The innermost layer of the new pixel detector is now closer to the beam pipe\footnote{The beam pipe has been replaced with a smaller diameter beam pipe in anticipation of the new detector during the long shutdown~1 in 2013-2014.},
and the fourth layer is also closer to the strip detector. This new geometry results in a better vertex resolution that is a prerequisite to be able to cope with the higher pileup at the LHC compared to previous years.

\begin{figure}
    \centering
    \includegraphics[width=0.32\textwidth]{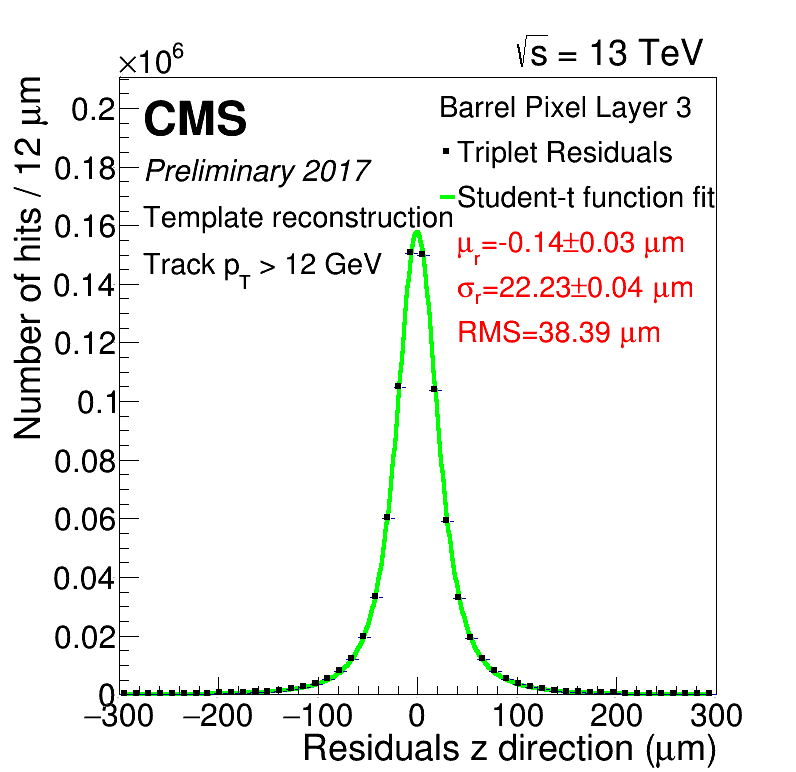}
    \includegraphics[width=0.32\textwidth]{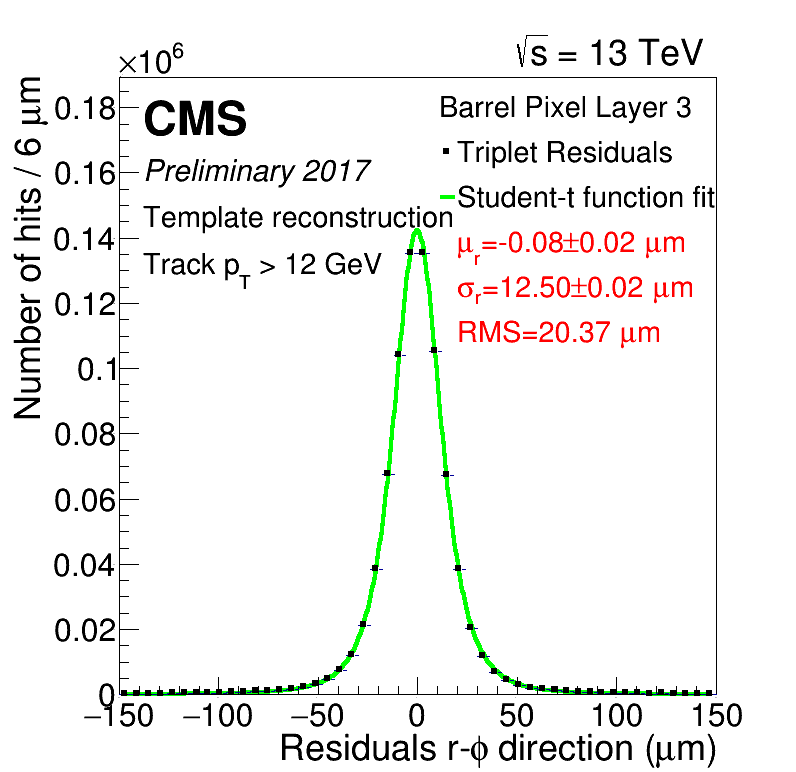}
    \includegraphics[width=0.32\textwidth]{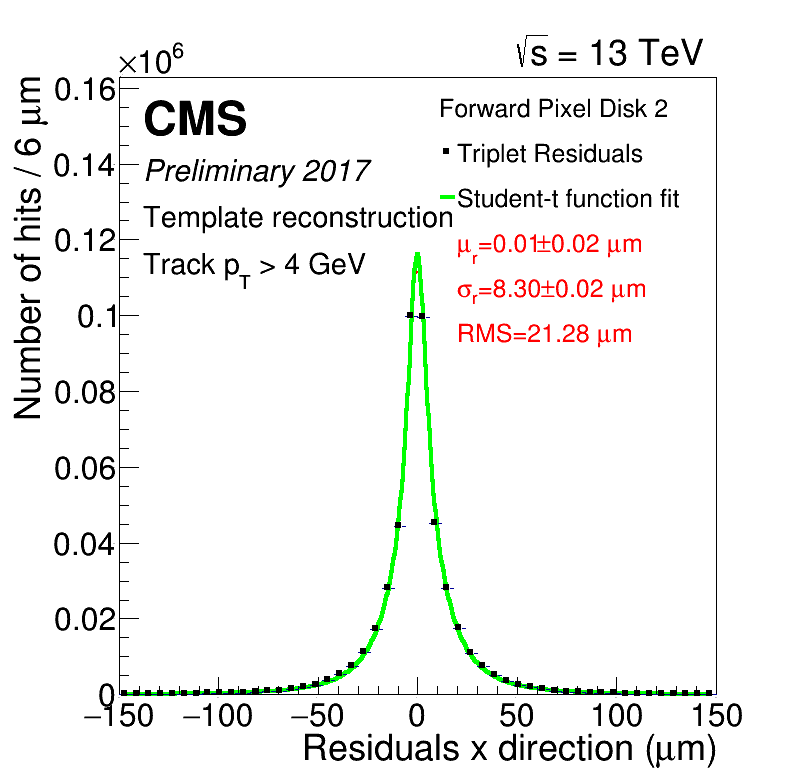}
    \caption{Residuals using template reconstruction in the barrel layer 3 and forward disk 2. Figures from \cite{cmstracker}.}
    \label{fig:res}
\end{figure}

One indicator of a well-calibrated, well-working pixel detector is hit resolution. The hit resolution in the pixel detector is determined using residuals, or differences between the cluster position of a hit and a well-reconstructed track with at least three hits in three layers/disks in the pixel detector. In this process, the position of the cluster is determined using parameters from a fit to simulation
called ``template reconstruction'' 
\cite{Swartz:2002kda},
 and the momentum of the original track is used to propagate the trajectory analytically from the remaining hit doublet. Simulations show \cite{trackerperformance} that all layers and disks have the expected resolution apart from layer 1. The latter suffers from higher than expected thresholds, reduced charge collection efficiency, double column inefficiency, and not yet perfectly calibrated measurements of cluster positions
\cite{Veszpremi:2017yvj},
which shows that the commissioning of this part of the detector is still ongoing.
 Residuals for disk 2 and layer 3 are shown in Fig.~\ref{fig:res}.

\section{Phase~1 pixel detector modules}
\begin{figure}
    \centering
    \includegraphics[width=0.48\textwidth]{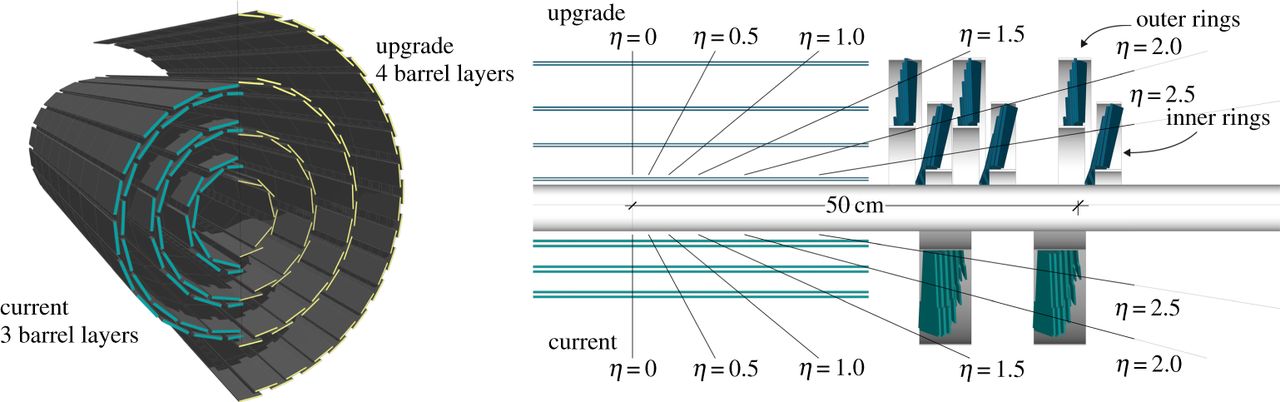}
    \includegraphics[width=0.48\textwidth, trim=480 500 50 80, clip]{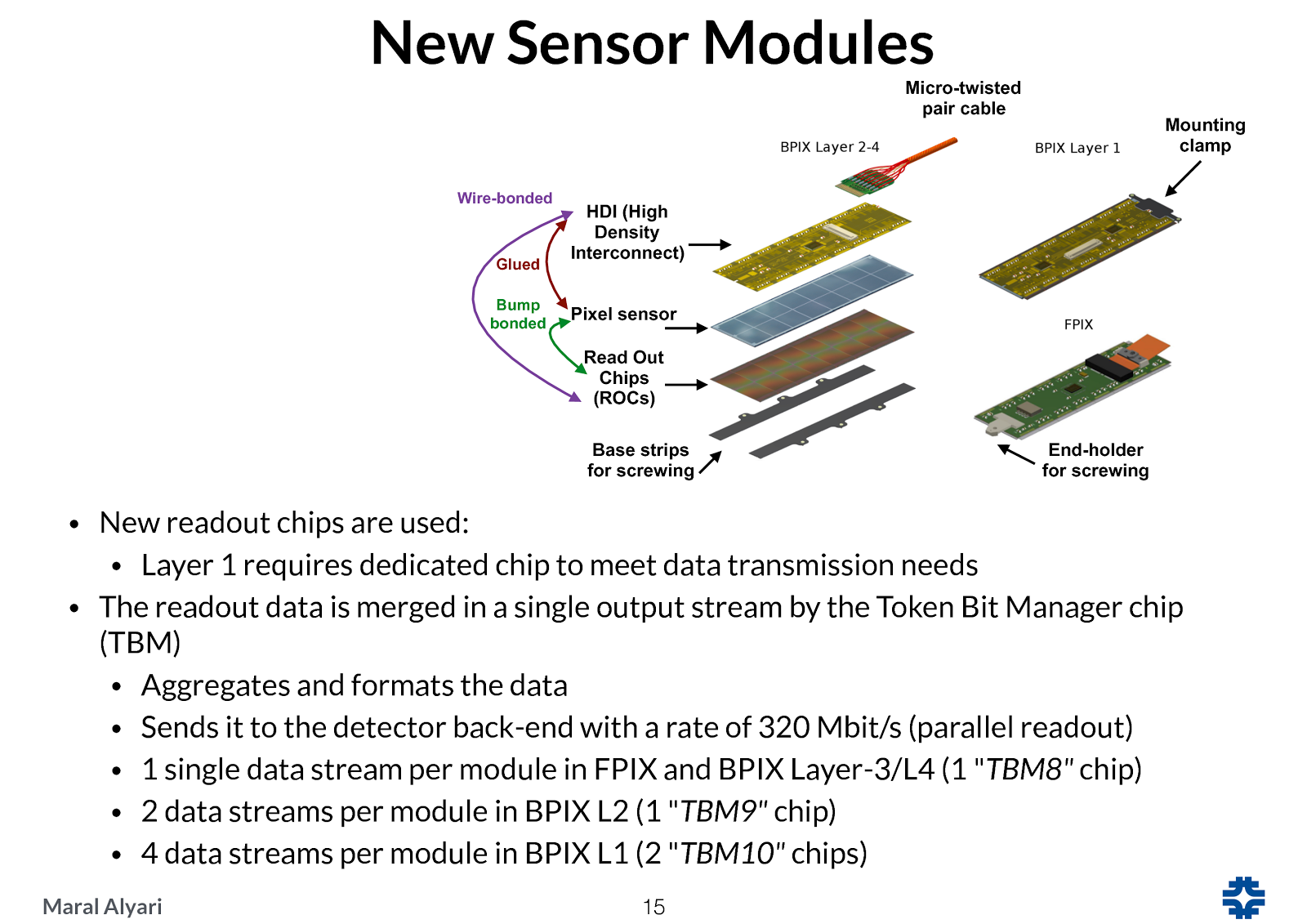}
    \caption{Geometry and modules of the phase~1 pixel detector. Figures from \cite{CMS:2012sda}.}
    \label{fig:layers}
\end{figure}

The innermost layer of the pixel detector is now 29 mm away from the center of the beam pipe.
This was 43 mm in the previous detector; see Fig.~\ref{fig:layers}. For this reason and the resulting increase in hit rate in the first layer of the phase~1 detector, special readout chips (ROCs) are used in this first layer that differ from the other layers.
Modules contain 285 $\mu\mathrm{m}$ thick n+~in~n silicon sensors with 66560 pixels, where each pixel has a size of $100\times150~\mu\mathrm{m}^2$ as in the phase~0 detector. Each sensor is bump bonded to the $2 \times 8$ ROCs per module. The readout changed from analog to digital compared to the phase 0 detector, and the ROC readout rate increased from 40~MHz to 160~Mbit/s. The readout bandwidth of the modules increased by a factor of 2 (and 4 for the barrel layers 1 and 2), which went from 40~MHz in phase~0 to 320~Mbit/s in phase~1 (after multiplexing two 160~Mbit/s data streams).\footnote{
    Note that the analog readout in phase~0 had 7 address levels, whereas the digital readout in phase~1 has only 2 levels.
}
With an additional bit to ensure DC balance between the optical links, this results in a readout rate of 400~Mbit/s.
All ROCs are 250 nm complementary-symmetry metal–oxide–semiconductor (CMOS) application-specific integrated circuits (ASICs) of $80\times52$ pixels with pulse-height readout and an 8-bit analog-to-digital converter (ADC).

\paragraph{Token bit manager}
The token bit manager
\cite{Bartz:2002rr}, or TBM, manages the 320~Mbit/s parallel readout at 400~MHz
\cite{KaStli:2013vja}.
Each layer 1 module is equipped with two TBMs with each two cores, that receive and send triggers to read out buffered data from the 16 ROCs that each module contains. Data are sent out over four streams to deal with the higher rates of the first layer. All other modules have one TBM, which on layer 2 has two data streams, and one data stream elsewhere. The data streams from two ROC banks of 8 ROCs each are merged inside the TBM.

\paragraph{Layer 2-4 and endcap ROCs}
The layer 2-4 and forward disk modules are mounted with the PSI46digv2 %.1r (PSI46digv2 for short)
ROCs
\cite{KaStli:2013vja,Kastli:2005jj}
that have a double column drain architecture.
Each ROC has a 4160 pixel array arranged in 80 rows and 52 columns that are organized in 26 double columns. Each double column is interfaced with a data buffer of 80 cells (this was 32 in the phase~0 detector) and a time stamp buffer of 24 cells (which was 12 in the phase~0 detector).
Each ROC also has a control interface block with readout logic, digital to analog converter (DAC) registers, inter-integrated circuit (I$^2$C) interface, ROC readout buffer, and the 8-bit ADC converter.
When a trigger arrives, double columns in the ROC with hits matching the trigger bunch crossing are drained, and data taking in those double columns stops until the digitized data is transferred to the readout buffer. As long as this readout buffer is not filled up, no additional dead time is introduced \cite{KaStli:2013vja}.

\paragraph{Layer 1 ROCs}
The pixel barrel layer 1 is mounted with modules containing ROCs of the type PROC600
\cite{Starodumov:2017exc}
 that are designed to withstand a particle rate of $600~\mathrm{~MHz}/\mathrm{cm}^2$.
Compared to the PSI46digv2, this new ROC can run the column drain continuously. % and does not require a buffer reset.
Additionally, each double column in the PROC600 has a data buffer with 56 buffer cells, where each cell stores a $2\times2$ pixel cluster address and
the four analog pulse heights through a dynamic cluster column drain mechanism. It also features a 40-cell deep time stamp buffer and can have seven pending column drains instead of three.
The expected dose for layer 1 over a collection of 300 fb$^{-1}$ at the LHC is 120 Mrad or 1.2 MGy, for which it was shown that the comparator thresholds and noise could be kept the same \cite{Starodumov:2017exc}.

\section{Phase~1 pixel detector services and data acquisition}
The phase~1 pixel detector has supply tubes to service the barrel detector and service cylinders for the end caps that provide low voltages to the digital and analog parts of the modules, high voltages for the sensor bias, and facilitate command transmission and I$^2$C programming as well as readout of the modules.
In the relatively short period of installation, it was not possible to replace all power cables; instead, DCDC converters are used to convert the supply voltages for the modules. By placing this and other service electronics outside the tracking volume, a reduced material budget was achieved.
\paragraph{DCDC converters} Since the pixel phase~1 detector has a factor of 1.9 more channels than the phase~0 detector, the corresponding current increase would be expected to result in a factor of 3.6 more power losses in the cables
\cite{Lipinski:2017noa}.
This problem was overcome by installing custom DCDC buck converters
\cite{Feld:2016uwj,Feld:2013pda}, or switched-mode power supply step-down converters,
based on the FEAST2 ASIC 
\cite{Faccio:2014iya}
that provide 3.6~V and 2.4~V of low voltage to the digital and analog parts of the pixel modules, respectively.
The existing power supplies were in turn modified to deliver 11~V to the DCDC converters.
The about 1200 DCDC converters are mounted in the barrel pixel and forward pixel service areas outside the active tracking volume at a pseudorapidity of about $\eta=4$.
They are controlled with a communication and control unit (CCU) \cite{Paillard:2002yn}.
%\footnote{
%As of writing these proceedings in November 2017, several percent of these DCDCs have started malfunctioning. The cause of the problem is still not understood and is actively researched.
%}.
The DCDC converters are cooled with bridges to the same CO$_2$ cooling pipes that also cool the modules.

\begin{figure}
    \centering
    \includegraphics[width=0.68\textwidth]{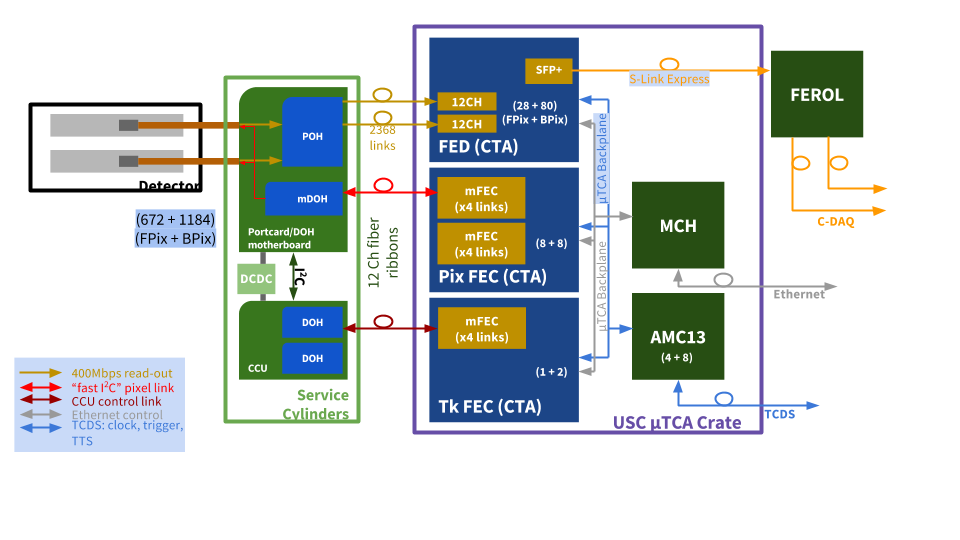}
    \caption{CMS pixel data acquisition and front end controlling. Figure from \cite{Kornmayer:2016lwv}.}
    \label{fig:pixeldaq}
\end{figure}

\paragraph{Data acquisition and front end control} An overview of the service electronics is given in Fig.~\ref{fig:pixeldaq}.
Most of the electronics in the forward disks is serviced by so-called portcards that contain optical hybrids used for module programming and readout.
Modules in the barrel detector are connected to connector boards through twisted pair cables. Digital signals are separated for layer 1 and 2, and layer 3 and 4, respectively.
ROC registers such as the regulated input voltages at the chips and comparator thresholds, as well as various signal delays, are programmed 
%The digital to analog converter registers\footnote{
%Examples of such registers are delays in the TBM and the comparator threshold (\texttt{VcThr}), analog and digital input voltage (\texttt{Vana},\texttt{Vdd}), and the number of bunch crossings back an event is to be read out on receiving a trigger (\texttt{WBC}) in the ROCs.
%} of the ROCs and TBMs are are programmed
with a front-end controller (FEC) through a digital opto-hybrid (DOH) via a bidirectional optical link.
Sixteen such so-called pixel FECs (``Pix FECs'') distribute a 40~MHz clock, trigger, and fast signals to the pixel modules
\cite{Kornmayer:2016lwv}.
Three so-called tracker FECs (``Tk FECs'') program auxiliary components in pixel supply electronics, such as opto-components and DCDC converters via an I$^2$C interface and peripheral interface adapter (PIA) port of a CCU.
 A pixel opto-hybrid (POH)
\cite{Troska:2012zz}
converts the electrical readout signal from the modules to optical data that are then sent to one of the 108 front end drivers (FEDs), which in turn decode the incoming data stream from the detector front-end. The FEDs each assemble all 24 channels’ (12 fibers) data into event fragments, and subsequently push these data to the CMS central data acquisition.
Both FEDs and FECs are Advanced Mezzanine Cards (AMCs) based on a CTA (CMS Tracker AMC) card
\cite{Hazen:2013rma}
 that is a variant of the FC7 card\footnote{
The FC7 card, which is a full size, double width $\mu$TCA AMC, is used by other CMS subsystems.
%, among others the CMS Trigger Control and Distribution System (TCDS), the
%CMS-TOTEM Precision Proton Spectrometer (CTPPS), the CMS Hadron Calorimeter, the CMS level 1 trigger, and the CMS muon track finder.
}
\cite{Pesaresi:2015qfa}
 which holds a Xilinx Kintex 7 Field-Programmable Gate Array (FPGA) and is capable to drive and receive links of up to 10 Gb/s.
Experience with a pilot system installed alongside the previous phase~0 detector and tests with the FED internal emulator helped in commissioning of the DAQ system
\cite{Akgun:2017vck}.

\section{Checkout and commissioning of the CMS phase~1 pixel detector}
\begin{figure}
    \centering
    \includegraphics[width=0.35\textwidth]{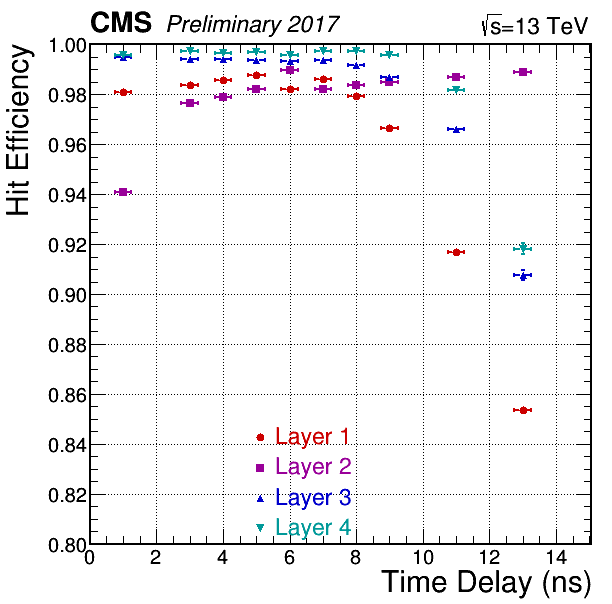}
    \includegraphics[width=0.35\textwidth]{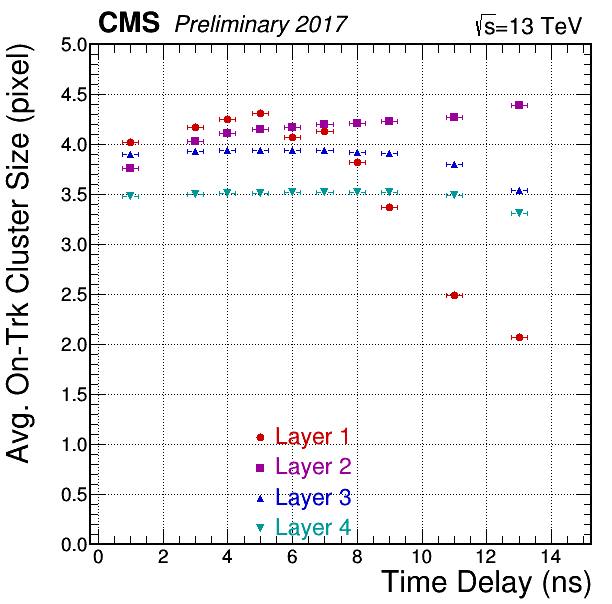}
    \caption{Hit efficiency and cluster size as a function of time delay for the barrel pixel layers. Figures from \cite{cmstracker}.}
    \label{fig:timing}
\end{figure}

\begin{figure}
    %and active channels of CMS in August 2017. Figure from \cite{cmstracker,activefraction}.}
    \centering
    \includegraphics[width=\textwidth]{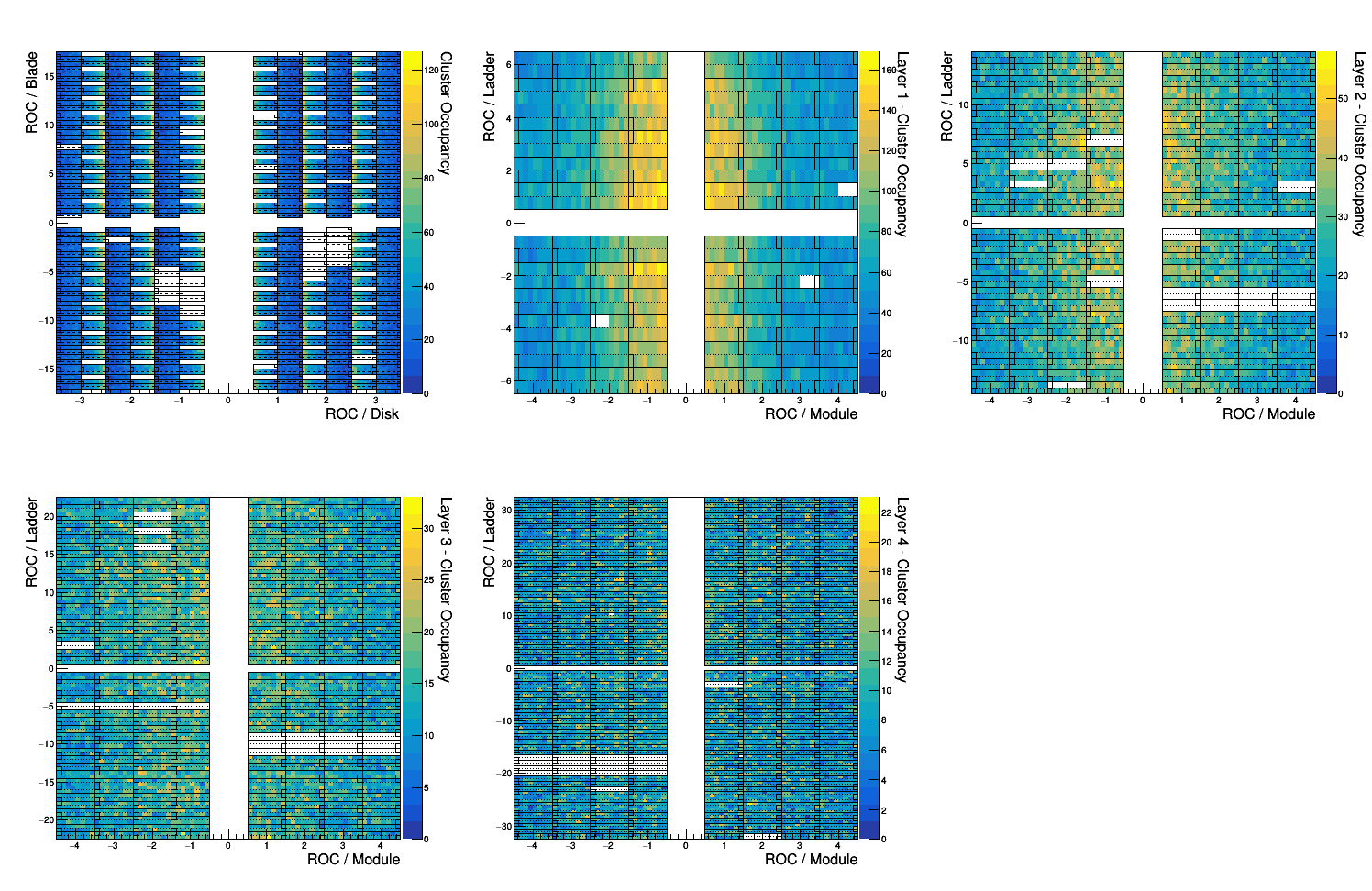}
    \caption{Occupancy map for the forward disks (x-axis) and overlaid disk parts (each part fanning outward as in a turbine) called blades (y-axis) for the endcap detector (top left), and the barrel layers 1-4 per ladder (y-axis) and module (1-4, x-axis, where 1 is lowest and 4 highest in $\eta$, respectively) in the phase~1 pixel detector \cite{cmstracker}. The white spots in layer 1 (top center) are examples of unresponsive TBMs following an SEU.}
    \label{fig:occupancy}
\end{figure}

%, even though the first collisions took place only on the 23rd of May\cite{cmslumi}.
After preparation of the power supplies, cooling, and optical fibers, the phase 1 detector was installed between February 28th and March 3rd, and the checkout took place in March and April. Around mid-April the pixel detector jointly took data with the other CMS subsystems for the first time.
%, and soon after the CMS solenoid reached 3.8 Tesla.
Notable challenges encountered during checkout and commissioning of the detector and current solutions to several problems are described below.
\paragraph{Detector timing}
Digital readout and transmission from front end to back end needs well-adjusted synchronization of signals, both internally and externally with LHC collisions.
The number of particle hits in a triggered collision and sensor efficiency are maximized by adjusting the relative phase difference between the internal clock of the ROCs and the bunch crossing of the event provided by the trigger in terms of LHC clock periods of 25 ns.
Signal propagation time is determined by the module cable length, position of the module connector on the connector board, and the routing distance to the DOH mother board.
Clock phases are programmable by chips on the DOH motherboard.
Since phases are not adjustable per single ROC or even single layer, phase and timing adjustment continues today to achieve higher efficiencies as well as improved cluster properties such as cluster size and charge \cite{cmstracker, viktorvertex2017}.
\paragraph{Timing of the PROC600 and PSI46dig}
There is a shift of half a clock cycle (12.5 ns) between the barrel layer 1 (PROC600) and layer 2 (PSI46digv2) chip, but their delay chip is shared in a single $\phi$ sector. The PROC600 is faster than PSI46dig.
The resulting discrepancy in layer 1 and layer 2 timing can be seen in Fig.~\ref{fig:timing}, where the hit efficiency is shown as a function of time delay.
The current solution is to speed up layer 2 and slow down layer 1 with a working point of 98\% efficiency for both layers.
One way of achieving this was to increase the layer 2 sensor bias voltage from 100 to 250~V which allowed it to gain 1-2 ns, as this results in faster charge collection.
%saturation of the charge thresholds.
%Settings that were not used were \texttt{ViColOr}, which determines above which voltage as a result of any pixel hits a double column is read out, and \texttt{Vana}, which sets the output analog voltage at the chip, to slow down layer 1 as the gain was small.
%The result is an efficiency overlap of 2-4 ns but with non-optimal cluster charge and size
%\cite{Veszpremi:2017yvj}.
%There is also half a clock shift layer 3 and 4 in one barrel $\phi$ sector, probably as a result of a faulty gatekeeper: using a different number of bunch crossings (\texttt{WBC}) in this sector solves this problem.
\paragraph{Thresholds}
A comparator threshold can be set for each ROC to determine above which charge accumulation a signal is sent out.
Higher comparator thresholds can result in pixels at cluster edges not being able to collect enough charge to pass the threshold, which in turn impacts cluster size and consequently hit resolution.
Low thresholds are also desirable as lower charge collection occur with more radiation damage.
It was possible to obtain low thresholds for layer 2-4 and endcap ROCs, which were about 1800 e$^-$
\cite{cmspixel} in September 2017.
The thresholds on the PROC600 modules are about 2500 e$^{-}$, which is higher than anticipated as a result of crosstalk \cite{Veszpremi:2017yvj}. Typical noise is of the order of 100 electrons for both chip versions.
%Several initial tests were done for the checkout of the detector.
%In order to optimize and test basic optical connection it was tested whether light comes from the fiber.
%The supply tube settings such as signal delays and light yield were adjusted until the idle pattern of a module could be seen.
%The external 400MHz TBM phase was adjusted so that TBM headers and trailers could be seen (no token passed), and the internal 160MHz TBM phases (ROC ports) were adjusted so that ROC headers could be seen.
%Finally, pixels were tested for hits by injecting and reading out test hits.
%The DAQ was tested with a FED hit emulator.
\paragraph{Active channels}
% compare phase~0
The fraction of channels included in data acquisition of the pixel detector was 95.5\%  mid-August
\cite{activefraction}.
By this time, the LHC delivered and CMS recorded already half the integrated luminosity of what was delivered and recorded in 2016 (almost 20 fb$^{-1}$ for CMS), respectively.
An occupancy map of the CMS pixel detector in August 2017 is shown in Fig.~\ref{fig:occupancy}.
A loose power supply line on the detector side results in no data in one of 32 sectors of the barrel pixel detector in layers 2 and 3 (top right, bottom left). A loss of a portcard in the forward detector causes a loss of sensitivity in that part of the detector.
One of the 32 sectors in the pixel barrel detector has the characteristic that layers 3 and 4 do not function above a certain temperature. To reduce the thermal load, the cooling temperature has been decreased before summer by 2$~^{\circ}\mathrm{C}$ down to -22$~^{\circ}\mathrm{C}$, and the fourth layer in this sector has been disabled.

\paragraph{Single event upsets}
Single event upsets (SEUs) cause several issues in the phase~1 pixel detector. The token bit manager electronics is not fully redundant, and sometimes stops sending tokens following an SEU. A TBM reset does not solve the problem; instead, the only way to reset the TBM is to powercycle it. This can be done through disabling and enabling DCDC converters that are attached to one or more modules. Since a TBM has two cores, an unresponsive TBM manifests itself as half a module without hits in the occupancy plots (or a quarter in case of layer 1); see also Fig.~\ref{fig:occupancy}.
SEUs also sometimes result in a loss of all modules connected to an entire readout group serviced by one portcard in the end caps. The solution here is to reprogram the portcard connected to this readout group to recover it. ROC SEUs are solved by reprogramming single ROCs.
In August~2017, downtime caused by the pixel detector was reduced with automatic disabling and enabling of DCDC converters connected to modules with TBMs affected by SEUs %\footnote{
%Because of the current malfunctioning DCDCs that tend to not recover after a powercycle, this automatic powercycling has been disabled since the beginning of October 2017 until this problem is understood.
%A minimum of manual powercycling is still done to recover stuck TBMs.
%}
and reprogramming of portcards.
For improved cluster properties in the barrel layer 1
\cite{Veszpremi:2017yvj},
 the pixel detector has a so-called ``private'' synchronized ROC reset where CMS triggers are paused and the pixel detector receives a reset, during which time other subsystems receive no such reset -- hence the name private.% The current cluster properties be seen in Fig.~\ref{fig:clusterprops}.
 %(1) CMS triggers are paused, (2) buffers are read out, (3) a pixel-only resync command is sent, whereafter (4) triggers are paused for another set of orbits before they are (5) reenabled. The current cluster properties be seen in figure \ref{fig:clusterprops}.
%\begin{figure}
%    \centering
%    \includegraphics[width=0.35\textwidth]{img/OnCluChargeNorm_Disks_TBMResetBGo14.png}
%    \includegraphics[width=0.35\textwidth]{img/OnCluChargeNorm_Layers_TBMResetBGo14.png}
%    \caption{On track cluster charge with ROC resets. Figures from \cite{cmstracker}.}
%    \label{fig:clusterprops}
%\end{figure}

\section{Radiation damage in the phase~1 pixel detector}
\begin{figure}
    %\caption{Leakage current in the in the pixel endcap detector (phase~0) and pixel barrrel (phase~0 and phase~1). Figure from \cite{cmspixel}.}
    \centering
    \includegraphics[width=0.53\textwidth, trim=170 412 200 268, clip, page=9]{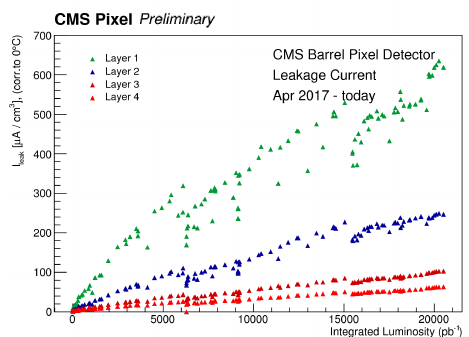}
    \caption{Average leakage current per layer in the pixel barrel detector normalized to 0$~^{\circ}\mathrm{C}$ from April 2017 until September 4, 2017.
    %Technical stops (TS), long shutdowns (LS), Year- of the End Technical Stops (YETS), and Machine Development (MD) periods include calibrations and increases in bias voltage, and can result in annealing and a decrease in leakage current.
    %Technical stops, long shutdowns, year- of the end technical stops, and machine development periods include calibrations and increases in bias voltage, and can result in annealing and a decrease in leakage current.
    Technical stops, long shutdowns, year- of the end technical stops, and machine development periods can result in annealing and a decrease in leakage current. The bias voltage was changed for layer 1 from 100V to 200V on August 23rd.
% was this really all the updates, that is, only http://cmsonline.cern.ch/cms-elog/1007009?
% the elog says: 
% New settings as of run 302055 LS 46.
%
%    L1: 350V
%    L2: 250V
%    L3: 200V
%    L4: 200V
%    R1: 300V
%    R2: 300V
%
%trip limit set to 1mA.
    Figure from \cite{cmspixel}.}
    \label{fig:leakage}
\end{figure}

In Fig.~\ref{fig:leakage}, the leakage current in the pixel barrel detector as a result of radiation damage is shown. %, in comparison to those for the phase~0 pixel detector.
LHC fills from the beginning of 2017 data taking are employed.
Only fills with stable beams declared where high voltage is applied to the sensor bias are included.
Leakage currents are measured within 10 minutes from stable beam declaration, and the average current per ROC is normalized to the active sensor volume $V$ in $\mu\mathrm{A}/\mathrm{cm}^3$, with 
$V~=~0.81~\mathrm{~cm}~\times~0.81~\mathrm{~cm}~\times~285~\mu\mathrm{m}$.
%$V~=~8.1~\times~8.1~\times~0.285\mathrm{~mm^3}$.
The currents are corrected to $0~^{\circ}\mathrm{C}$ according to the guidelines of the Inter-Experiment Working Group on Radiation Damage in Silicon Detectors
\cite{gibson:1483028,sze1981physics}:
\begin{equation}
I(T_{\mathrm{ref}}) = I(T) \left(\frac{T_{\mathrm{ref}}}{T}\right)^2\exp\left(-\frac{\mathrm{E}_g}{2\mathrm{k_B}}\left[\frac{1}{T_{\mathrm{ref}}} - \frac{1}{T}\right]\right).
\label{eqn:raddamage}
\end{equation}
Here $\mathrm{E}_{\mathrm{g}} = 1.21$~eV and $T_{\mathrm{ref}} = 273.15$~K.
In accordance with observations of temperatures of cooled phase-0 modules in the lab, the actual sensor temperature is taken to be 10$~^{\circ}\mathrm{C}$ more than the coolant temperature (which was -20$~^{\circ}\mathrm{C}$ and later -22$~^{\circ}\mathrm{C}$):
$T = T_{\mathrm{coolant}} + 10~^{\circ}\mathrm{C}$.
Temperature fluctuations are not taken into account. Currents are averaged over all high voltage channels in a layer. % and are shown per unit volume.
The overall rise in leakage current is as expected and comparable to that for the phase~0 detector. Periods of no beam result in annealing and a decrease in leakage current.

\section{Status of the CMS phase~1 pixel detector}
The phase~1 pixel upgrade detector was successfully installed during the winter 2016-2017 EYETS: this was a significant milestone in the CMS phase~1 upgrade project.
The 2017 pixel commissioning was challenging, but performance benefits are starting to be realized: the phase~1 pixel system is now successfully taking data, its data acquisition is performing smoothly, and initial studies show that performance of more complex functions like vertex finding is already better than with the phase~0 pixel detector \cite{viktorvertex2017}.
After successful commissioning of the phase~1 pixel detector, efforts are being undertaken to further improve the performance of the system.
%Even after the successful commissioning of the CMS phase~1 pixel detector to achieve performance of the same level as the phase~0 pixel detector under higher instantaneous luminosity however, commissioning will continue to improve performance%\footnote{
%As mentioned in an earlier footnote, at the time of writing the CMS pixel detector suffers from malfunctioning DCDC converters. An extra extraction will take place in the EYETS 2017/2018 (which is prolonged for this purpose) to study the origin of this problem and replace the malfunctioning converters.
%}.

\end{document}